\documentstyle[sprocl,epsfig]{article}

\def\be{\begin{equation}}
\def\ee{\end{equation}}
\def\bea{\begin{eqnarray}}
\def\eea{\end{eqnarray}}

\begin{document}

\title{HADRONIC PHOTON INTERACTIONS AT HIGH ENERGIES
\footnote{Talk presented by R.~Engel at the
XXVI Int. Symposium on Multiparticle Dynamics, Faro, Portugal, 1996}
}
\author{F.W.~Bopp}

\address{Universit\"at Siegen,
Fachbereich Physik, D--57068 Siegen, Germany}

\author{R.~Engel}

\address{Institut f\"ur Theoretische Physik\\
Universit\"at Leipzig, D--04109 Leipzig, Germany\\
and Universit\"at Siegen,
Fachbereich Physik, D--57068 Siegen, Germany }

\author{A.~Rostovtsev}

\address{ LPNHE Paris Universit\'e Paris VI, IN2P3-CNRS, France\\
and Institute of Theoretical and Experimental Physics
Moscow, Russia}


\maketitle

\vspace*{-8cm}
\begin{flushright}
Siegen University SI-96-12\\
October 1996
\end{flushright}
\vspace*{+7.5cm}

\abstracts{
A simple phenomenological introduction to the physics of multi-pomeron
exchange amplitudes in connection with the Abramovski-Gribov-Kancheli
(AGK) cutting rules is given. 
The AGK cutting rules are applied to obtain
qualitative and quantitative predictions on multiparticle production at
high energies. On this basis, particle production
in hadron-hadron scattering, photoproduction, and in particular
the transition to deep-inelastic scattering is discussed.
}


\section{Multi-pomeron exchange phenomenology}

Assuming that high virtual masses are damped due to the dynamics of the
the strong interaction, hadronic interactions can be described
by Gribov's Reggeon field theory (RFT)\cite{Gribov68,Gribov69,Baker76}.
In RFT,
hadronic scattering is characterized by the exchange of pomerons.
The total amplitude can be written as the sum of $n$-pomeron exchange
amplitudes $A^{(n)}(s,t)$.
For each $n$-pomeron graph one can define a theoretical "total" cross
section applying the optical theorem ($\sigma_{\rm tot} 
= \Im m \left( A\right)$, implicitly $t=0$ is taken)
\begin{equation}
\sigma^{(n)} = (-1)^{n+1} \Im m \left( A^{(n)} \right),
\hspace*{1cm} \sigma_{\rm tot} = \sum_{n=1}^{\infty} (-1)^{n+1}
\sigma^{(n)}\ .
\end{equation}
Here, an alternating sign has been introduced by definition to keep all
partial cross sections $\sigma^{(n)}$ positive.

\begin{figure}[!htb] \centering
\hspace*{0.25cm}
\psfig{file=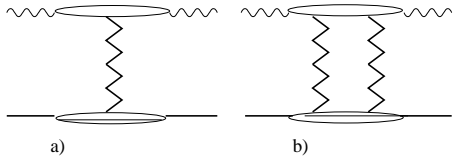,width=60mm}
\caption{
\label{allgra0} \em
Photon-proton scattering via pomeron exchange:
a) one-pomeron exchange graph, b) two-pomeron exchange graph.
}
\end{figure}

To keep the arguments as simple as possible, we restrict the
discussion to the graphs shown in Fig.~\ref{allgra0},
$\sigma^{(n)} \ll 4 \sigma^{(2)} < \sigma^{(1)}$ with $n > 2$.
Then, the total cross section reads 
$\sigma_{\rm tot} =  \sigma^{(1)} - \sigma^{(2)}$, where $\sigma^{(1)}$
and $\sigma^{(2)}$ are the cross sections of the one- and two-pomeron
exchange graphs, respectively.
In RFT, the energy-dependence of the cross sections $\sigma^{(1)}$ and
$\sigma^{(2)}$ can be estimated with
$\sigma^{(1)} \sim s^{\Delta_B}$ and $\sigma^{(2)} \sim
s^{2\Delta_B}$, $1+\Delta_B$ being the bare pomeron intercept.

\begin{figure}[!htb]
\centering
\hspace*{0.25cm}
\psfig{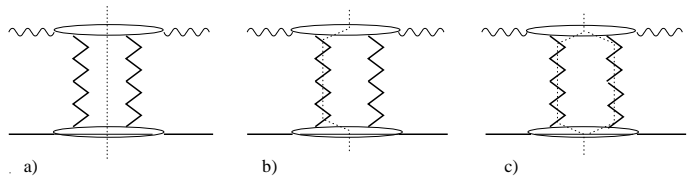}
\caption{
\label{pom2-cut} \em
Breakdown of the total discontinuity of the two-pomeron exchange graph
according to the AGK cutting rules:
a) the diffractive cut describing low-mass diffraction,
b) the one-pomeron cut, and
c) the two-pomeron cut.}
\end{figure}

To link particle production to elastic scattering amplitudes, we use
the optical theorem
together with the AGK cutting rules\cite{Abramovski73a}.
Up to trivial kinematical factors, the discontinuity of elastic
amplitude is equal to the squared matrix element describing particle
production.
In the framework of unitarity cuts, the discontinuity corresponds to an
unitarity cut moving the particles of all crossed propagators on mass
shell, e.g.\ all cut propagator lines become final state particles.
For example, let's consider the total discontinuity of the
two-pomeron graph. According to the AGK cutting rules, three different
cut configurations are giving the dominant contributions:
the diffractive cut with the weight -1
(Fig.~\ref{pom2-cut} a)), the one-pomeron cut with the weight 4
(Fig.~\ref{pom2-cut} b)),
and the two-pomeron cut with the weight -2 (Fig.~\ref{pom2-cut} c)).
The cross sections for the different final states of the two-pomeron
exchange graph are (i) diffractive cut: $\sigma_{\rm diff} = \sigma^{(2)}$,
(ii) one-pomeron cut: $\sigma_{1} = -4 \sigma^{(2)}$, and
(iii) two-pomeron cut: $\sigma_{2} = 2 \sigma^{(2)}$.
Note that the diffractive cut of the two-pomeron graph gets a negative
AGK weight, hence giving in total a positive, experimentally observable
cross section.
However, the cross section for the
one-pomeron cut of the two-pomeron graph is negative.
Since the one-pomeron cut of the
one-pomeron graph has the same inelastic final state as the
one-pomeron cut of the two-pomeron graph, one has to sum both
contributions giving together a positive cross section.

\section{Multiparticle production}

Concerning the topologies of the final state particles,
the total cross section is built up by the
the sum of the partial cross sections of the one- and two-pomeron
exchange graphs:
\begin{figure}[!htb]
\centering
\hspace*{0.25cm}
\psfig{file=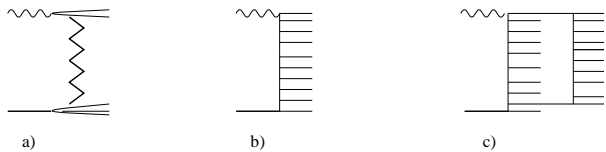,width=80mm}
\caption{
\label{pom2-ine} \em
Inelastic final states resulting from 
a) the diffractive cut describing low-mass diffraction,
b) the one-pomeron cut, and
c) the two-pomeron cut.}
\end{figure}
(i) one-pomeron cut (Fig.\ref{pom2-ine}~a))
$\sigma_1 =\sigma^{(1)}-4\ \sigma^{(2)}$,
(ii) two-pomeron cut (Fig.\ref{pom2-ine}~b))
$\sigma_2 =2\ \sigma^{(2)}$, and
(iii) diffractive cut of the two-pomeron graph (Fig.\ref{pom2-ine}~c))
$\sigma_{\rm diff} = \sigma^{(2)}$.

\begin{figure}[!htb]
\begin{center}
\unitlength1mm
\begin{picture}(120,55)
\put(0,0){\psfig{file=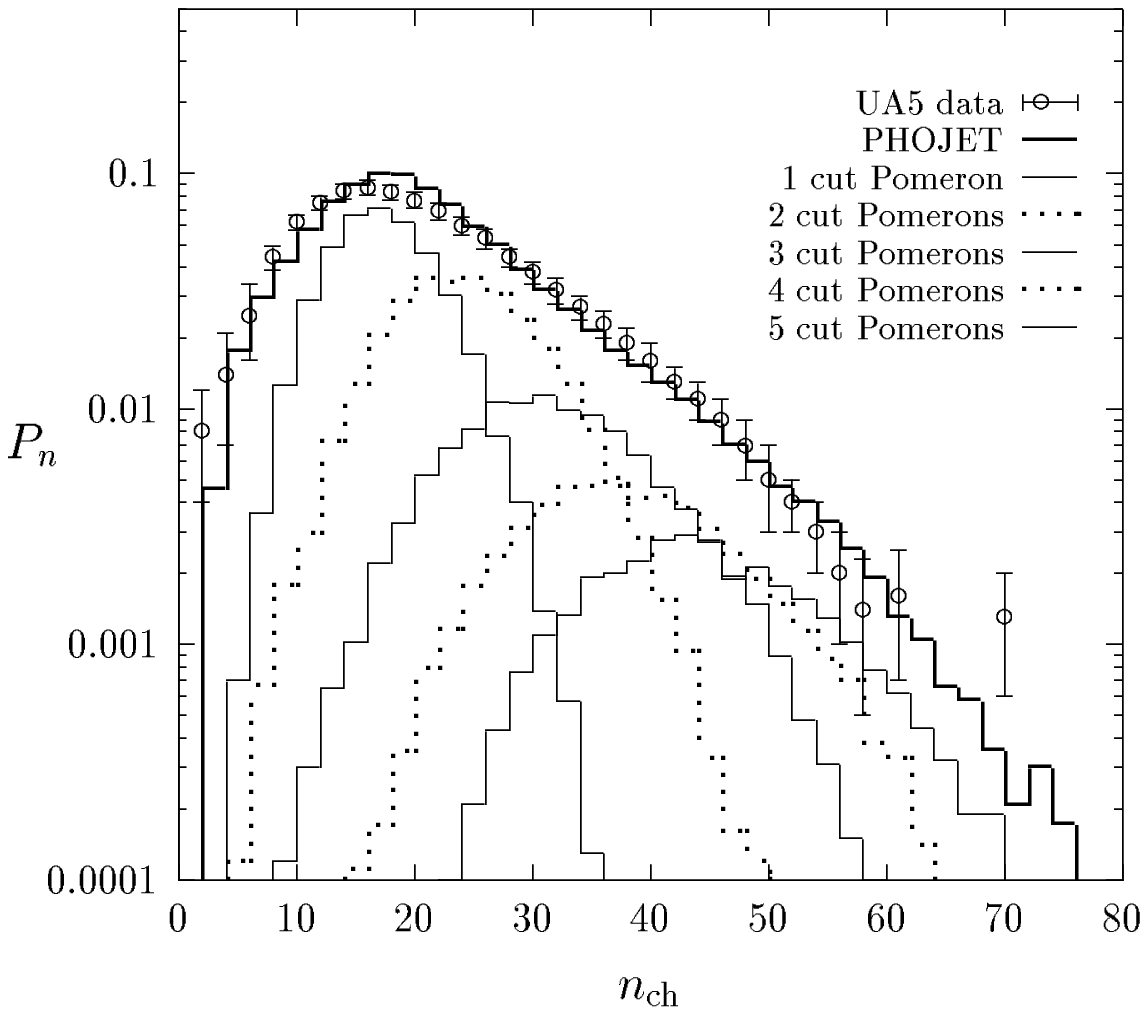,width=6cm}}
\put(65,0){\psfig{file=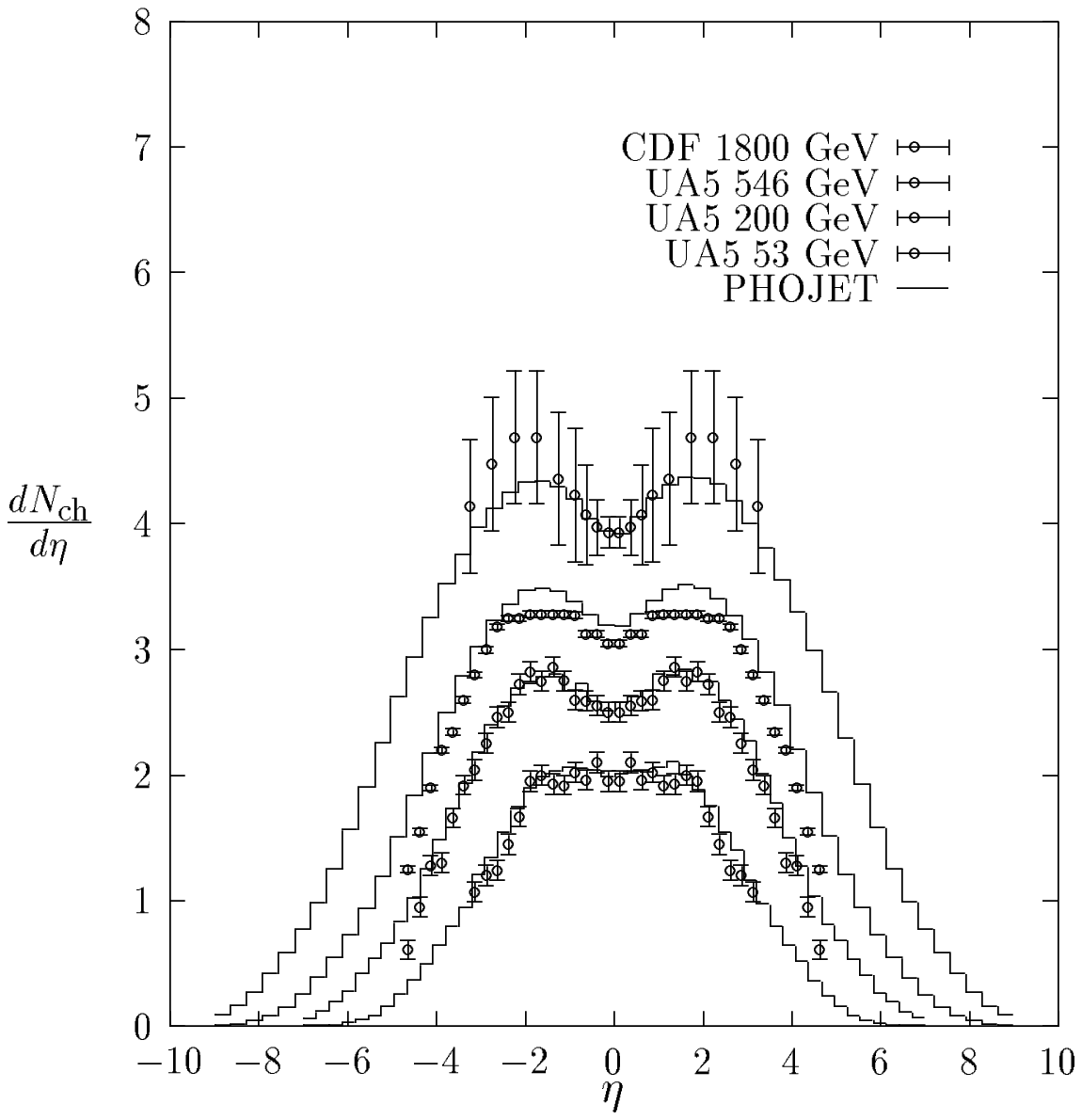,width=5.2cm}}
\end{picture}
\end{center}
\caption{\label{mulpom}
a) Decomposition of the multiplicity distribution in $p\protect\bar{p}$
collisions according to the number of generated pomeron cuts
at $\protect\sqrt s$ = 200 GeV. 
b) Increase of the charged particle pseudorapidity density in $p\bar p$
collisions with the energy.
}
\end{figure}

Let's consider a few applications of the AGK cutting rules to soft
multiparticle production.
The particle density in pseudorapidity of produced by a one-pomeron cut
is assumed to be almost energy independent (which is true for
longitudinal phase space models) and is denoted by $dN_1/d\eta$.
Assuming that a two-pomeron cut gives two times the particle
yield compared to the one-pomeron cut (in central pseudorapidity
region, see Fig.~\ref{pom2-ine}),
the inclusive inelastic charged particle cross section reads
\begin{equation}
\frac{d\sigma_{\rm ch}}{d\eta}\bigg|_{\eta\approx 0}
= 1\times \sigma_1 \frac{dN_1}{d\eta}
+ 2\times \sigma_2 \frac{dN_1}{d\eta}
+ 0\times \sigma_{\rm diff} \frac{dN_1}{d\eta}
= \sigma^{(1)}\ \frac{dN_1}{d\eta}
\label{AGK-cancellation}
\end{equation}
Due to the topology of diffractive final states, almost
no particles are produced in the central region in the case of a
diffractive cut.
Note that only the one-pomeron graph
determines the inclusive particle cross section in the central region
(AGK cancellation).
Then, the inclusive charged particle pseudorapidity density reads
\begin{equation}
\frac{dn_{\rm ch}}{d\eta}\bigg|_{\eta\approx 0} =
\frac{\sigma^{(1)}}{\sigma_{\rm tot}}\
\frac{dN_1}{d\eta}\bigg|_{\eta\approx 0}
\approx
\frac{\sigma^{(1)}}{\sigma^{(1)} - \sigma^{(2)}}\
\frac{dN_1}{d\eta}\bigg|_{\eta\approx 0}
\label{rho0-tot}
\end{equation}
Eq.~(\ref{rho0-tot})
allows to understand the observed behaviour of $\rho(0)=dn_{\rm ch}/d\eta$
in $pp$ and $p\bar p$ collisions.
With $\sigma^{(1)} \sim s^{\Delta_B}$ and $\sigma_{\rm tot} \sim s^{0.08}$,
one get a power-law increase of the central particle density.
This is confirmed by experiment\cite{Alner86b}: 
$\rho(0) \approx 0.74\ s^{0.105}$.
Since the two-pomeron graph has an cross section which
increases faster with the energy than the cross section of one-pomeron
graph, the model predicts also an increase
of the multiplicity fluctuations with increasing collision energies.
An consequence is the violation of KNO
scaling\cite{Koba72a,Koba72b} at high energies.
For example, model predictions for the
multiplicity distribution and the pseudorapidity distribution 
(calculated with the
{\sc Phojet} event generator\cite{Engel95a,Engel95d}) are compared with
data\cite{Ansorge89,Alpgard82a,Alner86b,Abe90} in Fig.~\ref{mulpom}. 
Furthermore, due to the characteristic structure of the one- and 
two-pomeron cut,
strong long-range correlations in pseudorapidity are naturally
explained. For a detailed discussion, see for example\cite{Capella94a}.

\section{Application to photon-hadron scattering}

All the important successes of the RFT and AGK based phenomenology 
encourage to apply this concept to photoproduction and the transition to
DIS. From $x \approx Q^2/(s_{\gamma^\star p} + Q^2)$ follows that low $x
= 10^{-3} \dots 10^{-4}$
and medium $Q^2 = 5 \dots 30$ GeV$^2$ correspond to the Regge limit of
$\gamma^\star p$ scattering.

There are two important new effects to note: {\bf (i)} the photon has a dual
nature and can interact as a gauge boson (pointlike) or a hadron
(resolved), and {\bf (ii)} the photon has an additional degree of freedom, the
photon virtuality.
Both effects give a handle to suppress
the relative size of the unitarity corrections (e.g.\ the relative size
of the multi-pomeron graphs compared to the one-pomeron exchange).
In direct photon interactions, there is no hadronic remnant to allow for
multiple interactions (e.g. two-pomeron exchange).
Considering $\gamma^\star p$ scattering with
not too small $Q^2$, the cross sections
$\sigma^{(n)}$, ($n \ge 2$)  are suppressed at least by a factor $1/Q^2$
compared to $\sigma^{(1)}$. Hence, multiple interaction contributions
decrease with increasing $Q^2$.
Furthermore, resolved photon interactions are more
suppressed with increasing $Q^2$ than direct photon interactions.


Concerning photoproduction, multiple interaction models predict 
a larger transverse energy deposit in resolved photon interactions 
than in direct photon interactions. In jet production via resolved photon 
collisions, additional interactions produce the so-called jet 
pedestal effect (which is absent in direct photon interactions).
Recently, this has been confirmed in jet measurements at 
HERA\cite{Abt93a,Aid95c,Derrick95b}.

The same arguments can be applied to
photon diffraction dissociation in photoproduction. 
In RFT inspired models, the diffractive cross section would grow with the 
energy like $\sigma_{\rm diff} \sim (s_{\gamma^\star p})^{2 \Delta_B}$. 
The measured
flat energy dependence is explained due to unitarity corrections: 
additional interactions produce particles filling the rapidity gap of the 
diffractive interaction. This can be parametrized effectively introducing
a {\it rapidity gap survival probability}\cite{Bjorken92a}.
The important point is that the rapidity
gap survival probability in hard diffraction differs significantly 
between direct and resolved photon interactions. 
Rapidity gap events with a resolved hard photon interaction are suppressed 
by a factor of about $2\dots 3$ compared to events with direct hard photon
interactions. Knowing the diffractive proton structure function (from DIS
measurements, gluon densities determined from scaling violation), 
this effect can be measured easily and allows to check the
concept of the rapidity gap survival probability.

For DIS, the limit $\sigma_{\rm tot} \gg \sigma_{\rm diff}$ 
can be satisfied by selecting events with not too small $Q^2$.
Then, AGK cancellation would predict for different $Q^2$ and
$x$, (using $\sigma_{\rm diff} = \sigma^{(2)}$, see Eq.~(\ref{rho0-tot}))
\begin{equation}
\left(\frac{\sigma_{\rm tot}}{
\sigma_{\rm tot} + \sigma_{\rm diff}} \right)
\frac{dn_{\rm ch}}{d\eta}\bigg|_{\eta\approx 0} \approx {\rm const},
\label{q2-independence}
\end{equation}
where the charged particle density is always normalized to the total
cross section. Note that the diffractive cross section $\sigma_{\rm
diff}$ is the cross section of 
quasi-elastic vector meson production and diffraction dissociation.
Furthermore, in this limit, the energy-dependence of the cross section
for quasi-elastic $\rho^0$ production $\sigma_{\rho^0}$
is almost the same as the energy-dependence of $\sigma^{(2)}$:
$\sigma_{\rho^0} \sim  \sigma^{(2)} 
\sim \left(s_{\gamma^\star p}\right)^{2 \Delta_B}$.

Further predictions are straight forward: for example,
with increasing photon virtuality,\\
(i) the jet pedestal effect decreases for jets at central c.m. 
pseudorapidity\\
(ii) multiplicity fluctuations decrease\\
(iii) long-range multiplicity fluctuations decrease.

Eventually, it should be mentioned that the phenomenology using
RFT and AGK cutting rules suffers corrections
due to energy-momentum conservation effects. It is very difficult to 
estimate these corrections without a detailed Monte Carlo model study.

\section*{Acknowledgments}

We acknowledge valuable discussions with A.~Capella and A.~Kaidalov.
One of the authors (R.E.) was supported by the Deutsche
Forschungsgemeinschaft under contract No.\ Schi 422/1-2.


\section*{References}


\begin{thebibliography}{10}

\bibitem{Gribov68}
V.~N. Gribov:
\newblock Zh.\ Eksp.\ Teor.\ Fiz.\ 26 (1968) 414

\bibitem{Gribov69}
V.~N. Gribov:
\newblock Zh.\ Eksp.\ Teor.\ Fiz.\ 26 (1969) 1306

\bibitem{Baker76}
M.~Baker and K.~A. Ter-Martirosyan:
\newblock Phys.\ Rep.\ 28C (1976) 1

\bibitem{Abramovski73a}
V.~A. Abramovski, V.~N. Gribov  and O.~V. Kancheli:
\newblock Sov.\ J.\ Nucl.\ Phys.\ 18 (1974) 308

\bibitem{Alner86b}
UA5 Collab.:  G.~J. Alner et~al.:
\newblock Z.\ Phys.\ C33 (1986) 1

\bibitem{Koba72a}
Z.~Koba, H.~Nielsen  and P.~Olesen:
\newblock Nucl.\ Phys.\ B40 (1972) 317

\bibitem{Koba72b}
Z.~Koba, H.~Nielsen  and P.~Olesen:
\newblock Nucl.\ Phys.\ C29 (1972) 201

\bibitem{Engel95a}
R.~Engel:
\newblock Z.\ Phys.\ C66 (1995) 203

\bibitem{Engel95d}
R.~Engel and J.~Ranft:
\newblock Phys.\ Rev.\ D54 (1996) 4244

\bibitem{Ansorge89}
UA5 Collab:  R.~E. Ansorge et~al.:
\newblock Z.\ Phys.\ C43 (1989) 357

\bibitem{Alpgard82a}
UA5 Collab.:  K.~Alpgard et~al.:
\newblock Phys.\ Lett.\ 112B (1982) 183

\bibitem{Abe90}
CDF Collab.:  F.~Abe et~al.:
\newblock Phys.\ Rev.\ D41 (1990) 2330

\bibitem{Capella94a}
A.~Capella, U.~Sukhatme, C.~I. Tan  and J.~Tr\^an Thanh~V\^an:
\newblock Phys.\ Rep.\ 236 (1994) 227

\bibitem{Abt93a}
H1 Collab.:  I.~Abt et~al.:
\newblock Phys.\ Lett.\ B314 (1993) 436

\bibitem{Aid95c}
H1 Collab.:  S.~Aid et~al.:
\newblock Z.\ Phys.\ C70 (1995) 17

\bibitem{Derrick95b}
ZEUS Collab.:  M.~Derrick et~al.:
\newblock Phys.\ Lett.\ B342 (1995) 417

\bibitem{Bjorken92a}
J.~D. Bjorken:
\newblock Phys.\ Rev.\ D47 (1992) 101

\end{thebibliography}

\end{document}